# ENHANCING MOBILE LEARNING SECURITY

Shaibu Adekunle Shonola and Mike Joy

Department of Computer Science, University of Warwick, CV4 7AL Coventry
United Kingdom

## ABSTRACT

*Mobile devices have been playing vital roles in modern dayeducation delivery as students can access or download learning materials on their smartphones and tablets, they can also install educational apps and study anytime, anywhere. The need to provide adequate security forportable devices being used for learning cannot be underestimated. In this paper, we present a mobile security enhancement app, designed and developedfor Android smart mobile devices in order to promote security awareness among students. The app can alsoidentify major and the most significant security weaknesses, scan or check for vulnerabilities in m-learning devices and report any security threat.*

## 1. INTRODUCTION

There have been many security incidents reported when using mobile gadgets ever since these devices have become popular, most especially in open operating systems [1]. With increasing use of mobile devices and applications for storing or accessing personal and sensitive information, many users are not aware of the growing security threats in using these devices and many users are also not aware thatsome mobile apps are not so secure. Asmore people use smartphones and tablets for their educational and financial activities, the more attractive these devices and their applications become targets to attackers with mischievous intents. Recent security surveys have reported a rapid increase in the number of mobile threats and the growing sophistication of the attacks. More worrisome is that an open and popular platform such as Android provides a comfortable environment to exploit and propagate security attacks [2].

To prevent or limit such undesirable attacks, Android developersare integrating security mechanisms and features that allow protection of users from malicious apps. Developing an effective and usable security model which is suitable for small portable devices is not an easy challenge. In fact, while addressing many security issues, the Android security model itself has many shortcomings [3], some of which are being addressed by Android security extensions such as Yet Another Android Security Extension [2].

Another issue on mobile device and m-learning security is lack of awareness or negligence among users as many learners do not regard security as important until an issue arises [4]. For example, despite the fact that the most widely used methods of authentication on mobile devices are PINs and passwords, a number of studies indicate that many mobile users are either unaware of, or do not bother to use, these security features [5]. A survey of 297 mobile phone users reported by Clarke and Furnell in [6] found that 34% of the participants did not use any PIN or password security,notwithstanding the fact that these security measures do not offer the best protection features.

Therefore, there is an urgent need to promote security awareness and education among users most especially among the young generations who are learners in higher educational institutions. Students need to understand the security threats and possess necessary knowledge onsecurity, when using their mobile device for learning and other purposes [7-8]. There have been a number of efforts on promoting security education among students for smart mobile devices; however, more efforts are still required in improving this security awareness, in term of designing and developing security enhancement apps for m-learning users to raise security awareness, scan devices for vulnerabilities and report any potential threats in their devices. This paper serves this purpose and it is organised as follows: section two presents related work on m-learning security, section three considers m-learning security app design, section four discusses the implementation and app evaluation of such an app, and section five and six present the results of the evaluation, section seven gives further discussion on the results and section eight presents the conclusion

## 2. RELATED WORKS

Researchers are now focusing their attention on mobile security issues due to a sharp rise in the number of reported mobile operating systems vulnerabilities, particularly in the Android platform by publishing many of their work on inherent security issues in mobile devices and m-learning platforms. Cai et al. [9]emphasise threats and attacks on user privacy by sniffing the sensors in smartphones and tablets. The authorshave developed a threat model based on the use of sensors and have designed a mobile framework for a defense system. This framework consists of three modules: (i) policy engine, (ii) interceptor, (iii) user interaction. The policy engine is based on application monitoring and profiling without requiring much user intervention and they consideredseveral policies under this,such as white-listing, blacklisting and information-flow tracking. The interceptor is interposed between the application and the sensors, and/or between the application and the network, and it enforces the decision of the policy engine. When the policy prohibits an application from accessing the network, the interceptor simply denies such access. The user interaction is not a mandatory but a desirable component, since it simply notifies the user by asking for their decision. For each module, different mechanisms are explored and discussed by the authors, however no real implementation of their engine is presented.

Ruitenbeek et al. [10] study the propagation of smartphone viruses,in particular, the effects of multimedia messaging system (MMS) viruses that spread by sending infected messages to other devices and propose several response mechanisms to quantify the effectiveness of virus mitigation techniques. The authors present a virus model which"parameterized and represented a wide range of potential MMS virus behaviours and identified four MMS virus scenarios": in each scenario the virus on the phone sends MMS messages with an infected attachment file to other phones, which are selected from the contact list of the infected phone as well as dialing a random phone number. If a user accepts the infected attachment file, the virus is installed, the target phone becomes infected and under the control of the attacker. The evaluated response mechanisms for each of the four scenarios are: (i) scan of all MMS attachments in MMS gateways to detect viruses; (ii) awareness promotion through user education; (iii) device immunization using software patches;(iv) monitoring for anomalous behavior and (v) blacklist of phones that are suspected of infection.Their experimental results revealed that any response mechanism must be quick enough to react rapidly to spreadingviruses. While their work is quite convincing, the authors acknowledged that an optimal virus response strategy must be able to address many different types of virus behaviour and there is room for evaluation extensions to their work.

Cheng et al. [11] proposed a virus detection and alert system for smartphones that detects viruses by collecting activity information from the smartphones, and performs joint analysis to detect both single-device and system-wide abnormal behaviours. They use a proxy to offload the processing burden from resource-constrained smartphones and, when a potential virus is detected, the proxy sends targeted alerts to infected devices and a subset of the uninfected devices to prevent the spreading. The authors claimed that their implementation results show the system can effectively prevent wide-area virus outbreaks with affordable overhead. However, their implementation was based on only the Windows Mobile operating system which only accounted for a smallfraction of the numbers of mobile devices worldwide. Also there are many other security issues apart from viruses which affect mobile gadgets [12].

This paper presents a practical mobile technology based learning approach to enhance security perception with a focus on students' needs. The approach describes the development of a mobile security app, which covers fundamental and emerging security issues and threats in modular form. Since using smart devices is becoming an integral part of students' daily activities, this approach provides a convenient and effective way for students to enhance the security of their devices.

## 3. REQUIREMENTS, DESIGN AND FUNCTIONALITIES

In order to understand the current security problems affecting smartphones and tablets being used for m-learning in developing countries, particularly in Nigeria, we investigated threats, vulnerabilities and attacks specific to these devices and examined ways in finding solutions to them. In particular, we reviewed literature, journal publications, policy documents and carried out surveys in higher education institutions, focusing our attention on high-level attacks. Having identified some issues, our intervention is to design and develop a mobile app that will enhance the security of any device that is installed on, through promoting awareness, scanning devices for vulnerabilities and reporting threats.

### 3.1 The App Architecture

The three main objectives of the intervention app are: 1) the security enhancement app helps students to understand not only classical in-built security models and solutions, but also problems in emerging areas of mobile security; 2) the app scansor checks students' m-learning devices in order to identify possible vulnerability in such devices, and in case some security lapses are detected, it gives tips on how to resolve them; and 3) the app presents a report on the identified security threats to the users as well as recommendations on tackling the identified issues. With these objectives in mind, the app architecture which is based on our proposed security framework for mobile client [13], is designed as follows.

The app adopts an activity-list or modular structure that organizes all security issues into a sequence of self-contained modules; eachactivity-list focuses on a specific security area identified during our research study and those obtained from academicpublications which include lecture notes, survey questions and interviewsas well as case studies. There are tips and awareness on each m-learning security issue which are backed by current research. Following the tips comes asecurity scan or check facility to analyse the mobile device and its installed app for any infection and provide advice on how to deal with the issue.

The app is designed and developed based on the Android platform due to the following three reasons. Firstly, the Android platform dominated the world-wide smart phone market with 82.8% as at second quarter of 2015 according to data from the International Data Corporation (IDC) in [14] on Worldwide Quarterly Mobile Phone Tracker. Secondly, the platform is popular andopen-source having less restrictive market policy which makes it a prime target for malicious applications. Lastly, the Android platform, although it has big major backers such as Samsung and Google, is very affordable in comparison to other mobile computing platforms.

The architecture of the app is outlined in the system diagram in Figure 1 below. The diagram provides an activity-list of security issues on which awareness tips are given and vulnerability scans that can be related to each issue.

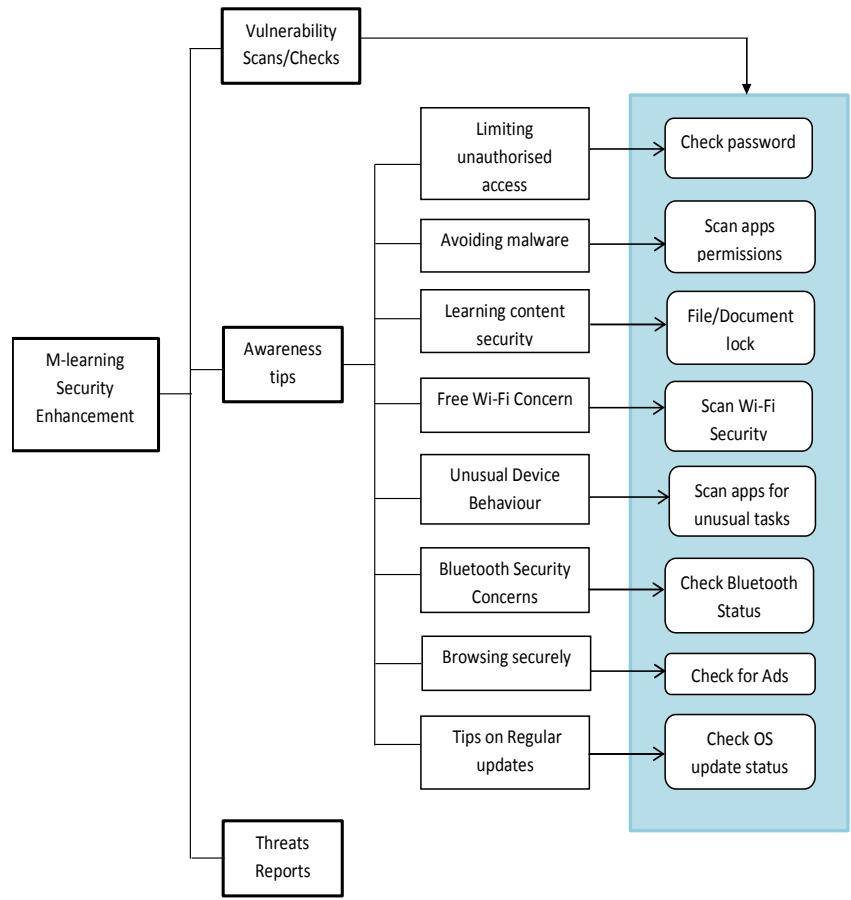

Figure 1: System diagram of the architecture of the app

## 3.2 Module Functionalities

The functionality of each module of the app is described below. The first functionality of the app is 'limiting unauthorised access' to the m-learning device by checking for password security. Our previous research indicated that many students do not have simple mobile device screen lock mechanisms such as Pattern, Pin or Password which may prevent unauthorized access to the device as well as the learning content in it. The learning documents stored on the device can be locked using file-lock password encryption in order to ensure their confidentiality and integrity if the device is lost or stolen. The second functionality is 'avoiding malware attack' by monitoring permissions requested by apps during installation and paying attention to unusual device behaviour. This functionality scans for permissions granted to all installed apps and any suspicious app can be triggered for further analysis and subsequent removal if necessary. The functionality also scans the device for unusual activities in term of resource usage (memory and processing) and gives notifications to the users.

The functionality on 'Free Wi-Fi Concern' warns the students on the implication of connecting to free Wi-Fi provided to the public by unknown people or organization. It scan for the security of Wi-Fi which the device is connected. The 'Bluetooth Security Concern' functionality advises students on the importance of ensuring that Bluetooth connection is disabled after use and checking the Bluetooth status of the device if it is currently switched off or no. As research shows that adware, spyware and malware spread through adverts that pop up when browsing the internet or from freely downloaded apps [15], the functionality of 'Browsing Securely' module checks the device if an advert blocker is already installed, and if not its provides a link to download and install an advert blocker. Regular checks are performed on the device to check if the latest updates have been installed for the OS. This is beneficial as most mobile device updates come with new security features and fixes to existing vulnerability in the software. The last functionality is the reporting section, which provides reports on all identified security threats in the device as well as some recommendations in fixing them.The flowchartin the Figure 2 below shows the basicflow of activities within the app.

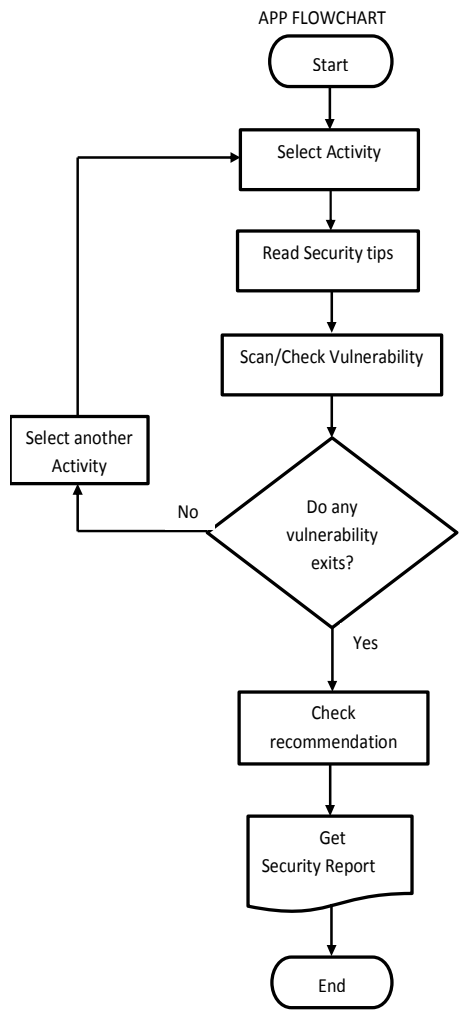

Figure 2: The app flowchart

## 4. DESIGN

The development and implementation of them-learning security enhancement app as an intervention is the main goal of this research. The content of this app reflects users' requirements which were gathered initial survey study [8]. The screen shots for home and activitypages of the app can be seen as shown below. From the home screen (Figure 3), users can move to the activity-list page by clicking on the "Security tips and scans" tab.

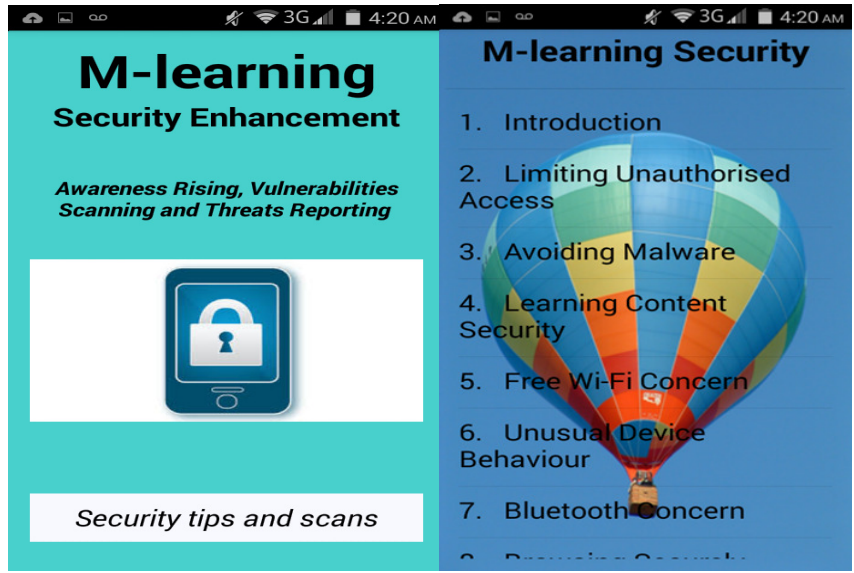

Figure 3: The app home page and activity list

From the activity-list page, users can select which security activity to explore in order to learn more about it and scan for vulnerabilities. For example, if a user wishes to detect unusual device behaviour, they can select the "unusual device behaviour" activity and start the scanner by clicking on the service button at the top on the page (Figure 4). This enables the app to start monitoring activities and sends notification if abnormal activity is detected.

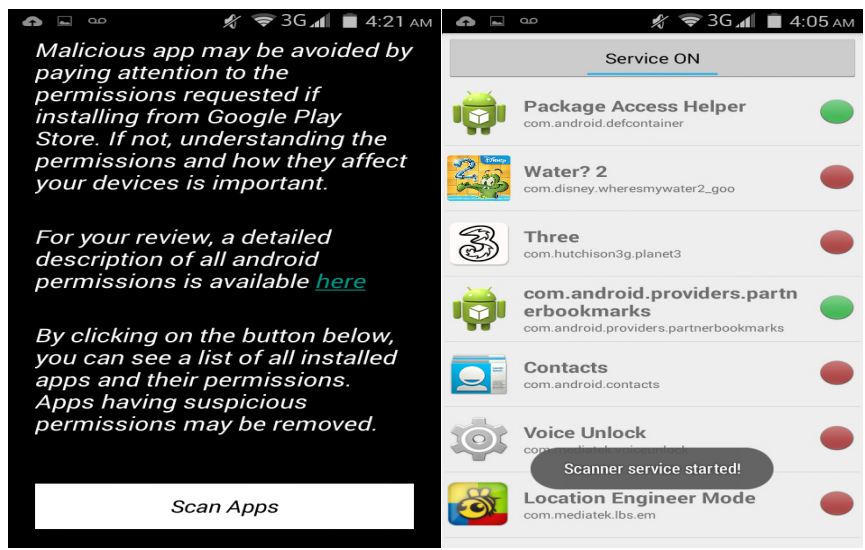

Figure 4: The app scan service

Similarly, a user can check the security of the Wi-Fi connected to by selecting "Free Wi-Fi concern" from the activity list and further select "Wi-Fi Security" (Figure 5)

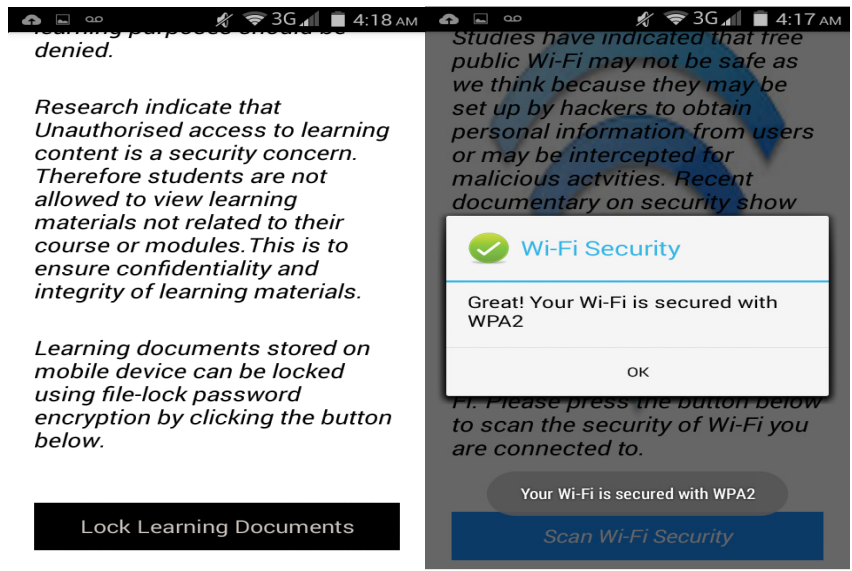

Figure 5: The app activity document lock/Wi-Fi security

The security reportpresents some findings on weaknesses in the m-learning device as well as makingappropriate recommendations to avoid further security breaches in future (Figure 6).

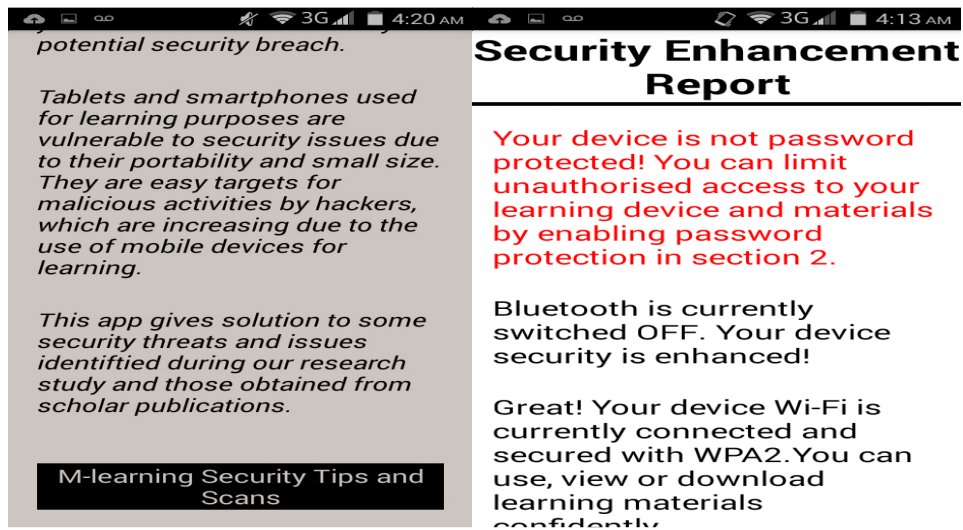

Figure 6: The app tips and enhancement report

## 5. APP IMPLEMENTATION AND EVALUATION

The first stage of the app implementation was a pilot stage where it was presented to postgraduate students for installation and use. Their initial assessment was positive, encouraging and valuable, their comments are incorporated into the actual released app for proper implementation. The main implementation stage involves distribution of the app to students and academic tutors in computer sciences in two Nigerian Universities for proper evaluation. The students who participated in the exercise were able to observe the app's features and usage. Theyweremade aware of the security

International Journal on Integrating Technology in Education (IJITE) Vol.5, No.3, September 2016

issues of their smart devices and learned some protection mechanism to reduce the risk of being exploited. Attributed to the convenience of m-learning and the daily use of their devices, students were more engaged in learning security and got insight of the security concepts through their hands-on practices and research studies cited in the app.

Evaluation took place after the app had been used for a couple of weeks in order to obtain feedback from the participants and to assess the app functionalities. As suggested by Miettine et al. [16]and Oyelere et al. [17],monitoring data in a mobile environment can be a challenge due to administrative, technical and conceptual limitations.While three methods are used for feedback collection and evaluation, only two methods are analysed in this paper. The firstis a questionnaire/feedback form which is attached to a section of the app. This is expected to be completed by all users within a period of two weeks after installation and first use. External link to the questionnaire can be provided if requested by the participants. The secondmethod for data collection is a set of interview with certain participants who are very conversant with our research work. The group include tutors in higher education institutions in Nigeria and colleagues at the University of Warwick.

### 5.1 Result from questionnaire

A survey was carried out among 110 students, most of whom were undergraduates in Nigerian Universities as shownin Table 1 below. Most of the participants are male between 20 to 25 years old.

Table 1. Gender/Age group demography

|  | Under 20 | 20 - 25 | 26 and over | Prefer not to say | Total |
|---|---|---|---|---|---|
| Female | 9 | 26 | 8 | 3 | 46 |
| Male | 6 | 37 | 16 | 5 | 64 |
| Total | 15 | 63 | 24 | 8 | 110 |

An initial question was about usefulness of the app. Almost all the participants said that their security knowledge improved by tips and information given by the app on how to keep their devices safe, as shown in Figure 7.

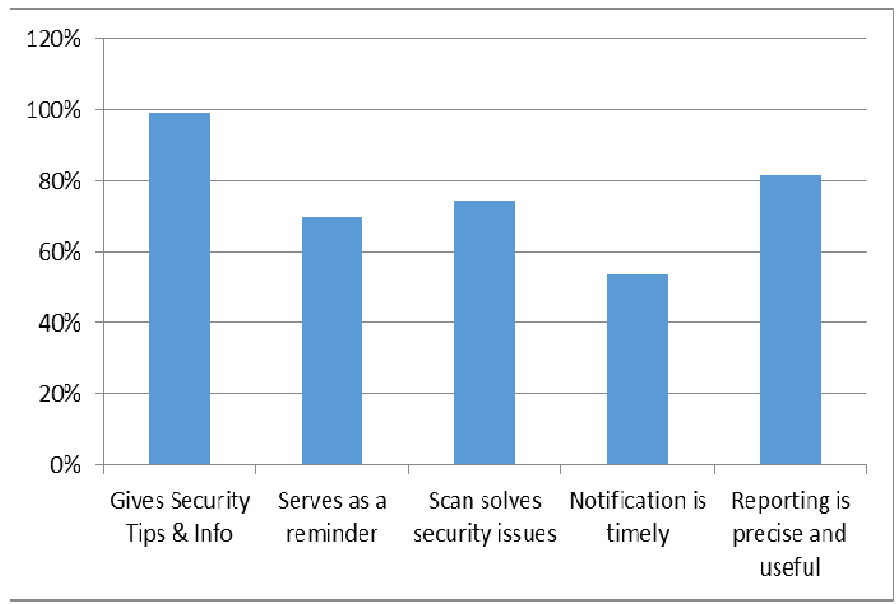

Figure 7: How the app is useful to the participants.

Another question was asked about the features of each section of the app. The rating of the features from the participants is shown in Figure 8 below. All the functionalities except the summary section have above 70% ratings from the participants.

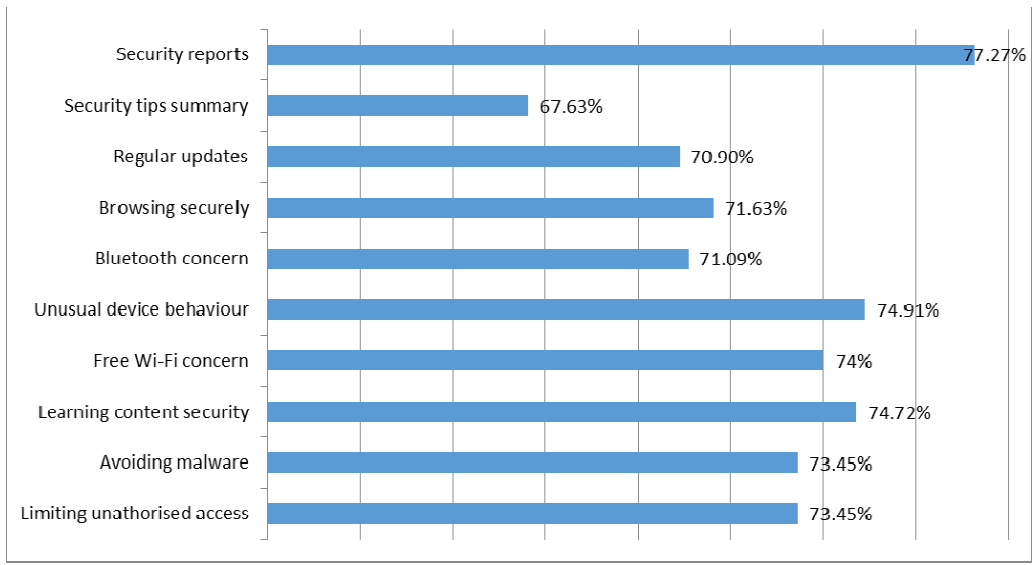

Figure 8: The features of the app

An interesting question from the survey also asked if the participant has perceived or experienced any security threat on their m-learning device before and if the app addressed the threat(s). 65.45% of the participants indicated that they had perceived/experienced threats before and 70.83% out of those participants said that the app addressed such security threats. An important feedback from the survey is if the security enhancement app meets the expectation of the participants and 77.27% responded 'Yes' while a meagre 2.73% responded 'No' and 20% indicated that 'They are not sure'. The last part of the questionnaire is to obtain the participants' general opinions on the functionalities built into the app in terms of fitness of purpose, enhancement of their security, improvement in their awareness, ease to download/install and contribution to their security knowledge. The responses are shown in Figure 9 below.

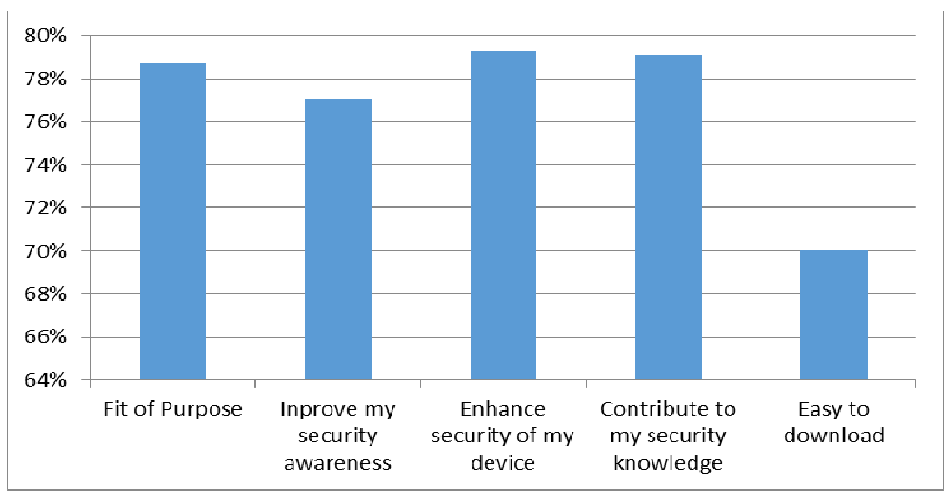

Figure 9: Users' opinions on the app functionalities

## 5.2 Results from Interview

An evaluation interview was conducted with 15 academic staff in Nigerian Universities. Table 2 shows the demographic information about the interviewees.

Table 2: Gender/Age Group of interview participants

|  | 30 - 39 | 40 - 49 | 50 - 59 | Total |
|---|---|---|---|---|
| Female | 2 | 2 | 0 | 4 |
| Male | 2 | 7 | 2 | 11 |
| Total | 4 | 9 | 2 | 15 |

During the interview, four main questions were asked about the features which have been built into the app, some of which are similar to those in the questionnaire. The first question is "which section of the app do you find very useful in terms of security?" and the response is plotted in Figure 10. Another question from the interview is, "which area does the app perform best in terms of awareness promotion, vulnerability scan and threat reporting?" Awareness promotion is considered the best with 66.67%, followed by vulnerability scan with 60% and lastly the report and alert with 53.33%

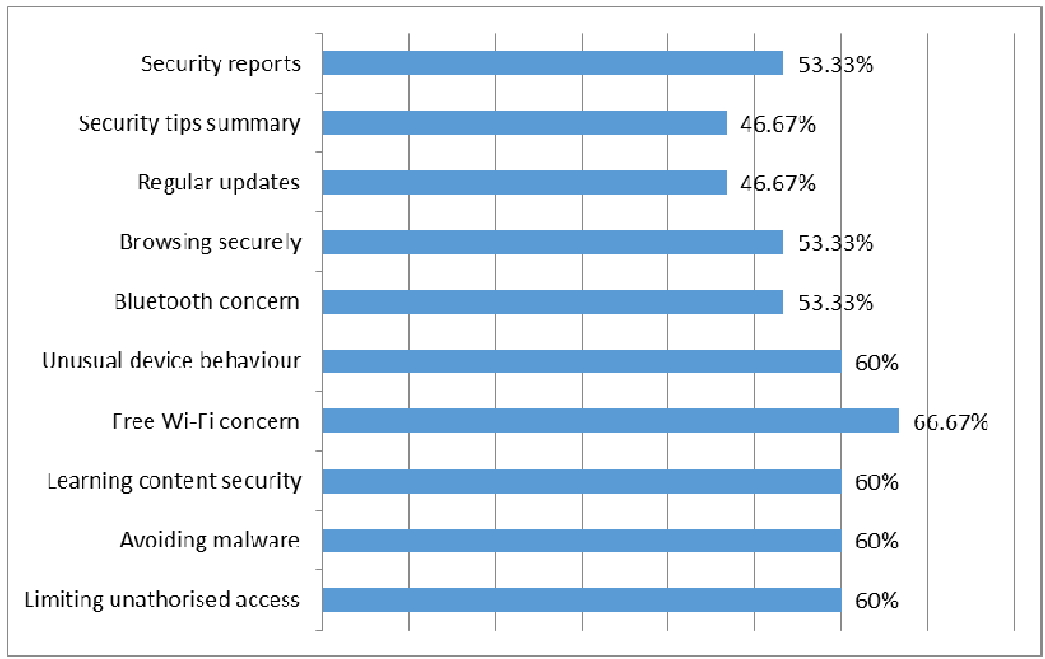

Figure 10: Users' opinions on the section of the app in terms of security

Many participants also believed that all the sections of the app are purposeful because they address some of the prevalent security issues that are common among students in Nigerian Universities. They also found all the sections of the app very useful in term of security diagnostics and reporting. Based on various benefits of the app with reasons mentioned earlier, all the interview participants thought that the app improves the security of their devices and the learning content. They all confirmed that the security enhancement app is fit for purpose.Further reasons given by the lecturer participants on why they think the app improves the security of their m-learning devices and why the app is fit for purpose are given in the conclusion section below.

International Journal on Integrating Technology in Education (IJITE) Vol.5, No.3, September 2016

## 6. COMPARING INTERVIEW AND QUESTIONNAIRE RESULTS

Due to similarity in some questions and feedback from questionnaires and interviews, it is necessary to compare them together in order to draw a logical correlation. First to be considered are the features of the app. Figure 11 below shows the findings between interview (lecturers) and questionnaire (students) feedbacks.

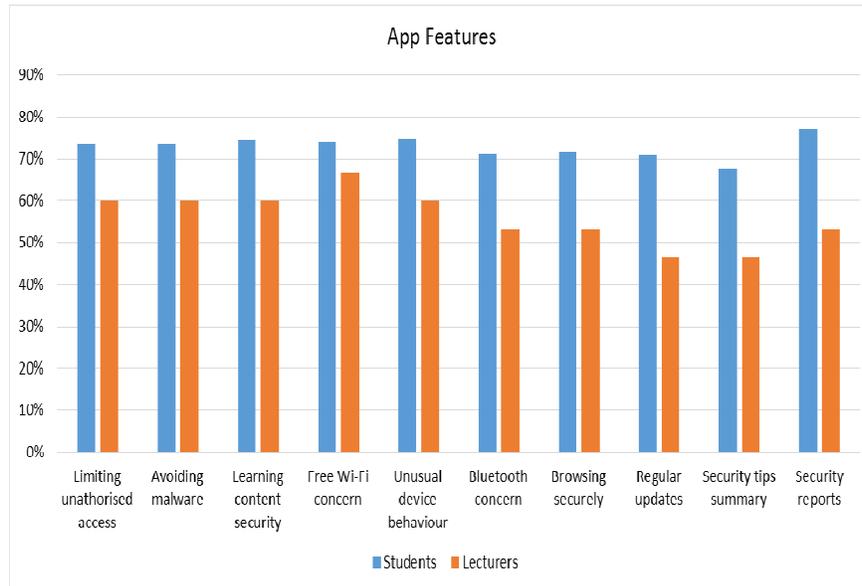

Figure 11:The app security features

It can be observed that the students rated all the app features higher than the lecturers. Are the students more generous and the lecturers more factual?

In comparing the app functionalities and purpose between the feedback from lecturers and students, the Figure 12 below says it all. While the students rated the all app functionalities higher than the lecturers, except for notifications and alerts which they rated the same, all the lecturers (100%) indicated that the security enhancement app is fit for purpose. Only 78.7% of the students share the same view.

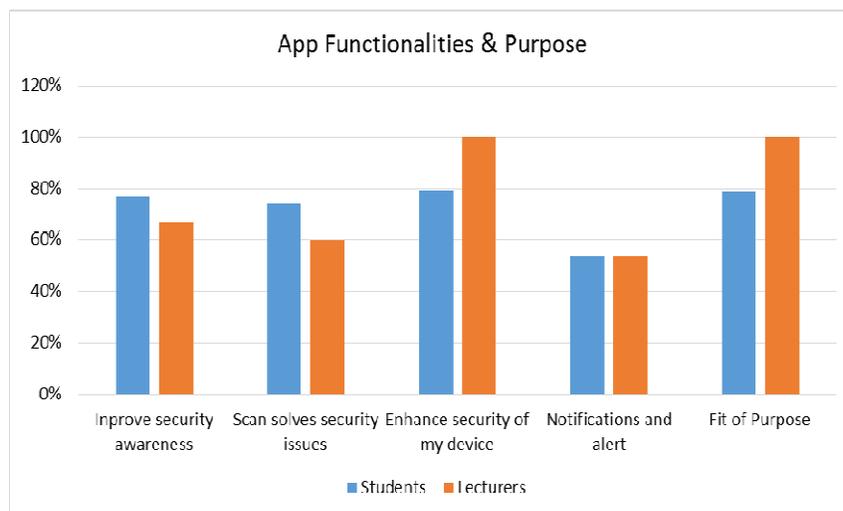

Figure 12: The app functionalities and purpose

## 7. DISCUSSION

Taking the security features of the app in turn, some respondents indicated that 'Limiting unauthorised access and learning content security' are useful to the students because of its file-lock and password mechanism which are related to some area in computer security syllabus. 60% of the participants said that they found 'Avoiding malware and unusual behaviour' helpful because they have had issues with malware before and would do anything to avoid it. Some participants also think it is beneficial because of the scanning functionality for vulnerability of the installed apps and the notification alerts, after starting the service scanner, are good. Two-third of the participants indicated that 'Free Wi-Fi Concern' is the most useful part because they can connect to any available free hotspot without thinking about security. The app feature is also educative for many students who always look for free Wi-Fi to connect to regardless of security implications. Just above half of the participants found the Bluetooth security feature interesting because they often forget to disable Bluetooth after use and also browsing securely section because they have had issues with advert pops up on our devices before. Information on how to block pop ups is a good idea to them.

Significantly, many of the participants believed that all the sections of the app are equally good and useful since they perform different activities which are necessary in providing adequate security for m-learning devices and that all the sections of the app are educative, informative and valuable. Regarding the area in which the app performs best in terms of awareness promotion, vulnerability scans and threat reporting, some of the academic participants indicated that the best module of the app is the awareness promotion because of the following reasons:

- It improves students' education on keeping their m-learning device.
- It complements their efforts in passing knowledge to the students because it enlightens the students about their mobile device security.
- The awareness on the danger of learning the Bluetooth on and as well as the research figure to support your claim is educative.
- The awareness promotion reminds the students that adware and spyware spread while downloading some free apps.
- From the tutors' discussion with students who used the app, some of them (the students) expressed that they have gained security knowledge through the use of the app.
- Some of the participants chose the vulnerability scan and check functionality because of the following reasons.
- Scanning app permissions is an interesting way to detect suspicious or malware infected app.
- The security scans/checks show the vulnerabilities and threats in their mobile device.
- The vulnerability scan performs best because it can detect potential security threats in apps and the device OS.
- Some are fascinated by the vulnerability scan as they use the check and scan to identify any security issue.
- The vulnerability scan is the most important part because it is where the actual security weakness identification and protection take place.
- One participant said, 'through the app, I enjoy testing the Wi-Fi connection anywhere I go by checking the Wi-Fi security'.
- Another participant said 'with the use of the app, my knowledge on adware and spyware has improved. I have also installed advert blocker as recommended, thus prevent advert pop ups on my screen unnecessarily'.

Many participants also indicated that the awareness tip and vulnerability scans are great features of the app because they serve as 'learn and practice' security sessions. Meanwhile many participants preferred the reporting section because the report identifies all the threats or security lapses in their device at once and provides suitable recommendations. The report also serves as a summary of all the results of the scans/checks features of the app. One participant observed that the report is interesting because the threats are highlighted in red colour, making him to pay more attention to them.

In summary, the evaluation study was successful because it gave us some feedback on what the students and educators think about our m-learning enhancement app, helped us determine whether they find the features devices useful and their opinion on the role of securing m-learning in education. It should, however, be pointed out that the sample size of this study limits generalization of the results; nevertheless, it does give a first glimpse on understanding the importance of m-learning security in higher education with Nigerian students.

## 8. RECOMMENDATIONS & CONCLUSION

There are many opinions which respondents made about the app, all of which are positive ones, the general opinion is that it is a good, simple, educative security app that is easy to use and understand by anyone who is interested in securing their sensitive information and learning contents, as they will find the app resourceful. Further, the app is excellent in reminding students about taking necessary precautions when using their devices and everyone who is security conscious will find the app impressive. Another opinion about the app that is worth mentioning is quoted as: 'I really like the app functionalities and I believe it is relevant in providing security services. My general opinion is that the app can be relied upon as a good security tool in protecting mobile devices'.

### 8.1 Recommendation

Despite the good rating, some recommendations and suggestions on improving the app are as follow:

- Modern biometric security features may be incorporated into the app such as finger prints and voice recognition instead of convectional file lockers and passwords mechanism
- Addition of more security notification alerts to other sections of the app aside the unusual behaviour section.
- To include prompt notification alert to Bluetooth and Wi-Fi sections if possible, rather than scanning fully before alert.
- Security issues on copyright materials can be included in future. That is copyrighted soft copy should not be shared without the author's permission, in form of DRM
- The app should distinguish real malware from other process and memory intensive app.
- The developers should keep updating the app in line with future security threats.

These recommendations will be considered in the future releases of the app.

### 8.2 Conclusion

We felt that the development of m-learning security enhancement app was necessary in order to raise students' awareness, augment existing security in m-learning devices and provide information on reducing threats. This article presents an enhancement app to provide security education and awareness among the students who engage their mobile devices for learning. The app helps in securing the learning contents on the portable devices through file-lock mechanism and gives students and teachers alike, the opportunity to practice simple security tasks. The security enhancement app does weakness checks or scans and offers appropriate recommendations. The monitoring facility of the app helps to monitor other apps which may be malware or spyware, through the scanner servicesand sends regular notifications to the users regarding any security issues or suspicious app. The app is considered fit for purpose because it helps to solve some of the security issues that students have encountered in the past.Above all, the app does what it says as it provides extra security facilities in addition to normal device security. Thus, the app enhances the in-built security features of mobile devices.

**Authors**

**Shaibu Adekunle Shonola** is a PhD student and researcher at the Computer Science Department, University of Warwick, UK. His research interests include mobile computing and security, educational technology, computer science education, adaptive and human centred systems.

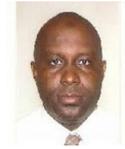

**Mike Joy** is an associate professor at the Computer Science Department, University of Warwick, UK. His research interests are in the fields of educational technology, computerscience education, agent-based systems and internet software.

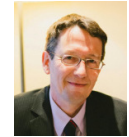